\documentclass[a4paper,twocolumn,journal, romanappendices]{IEEEtran}
\usepackage[T1]{fontenc}
\usepackage[latin9]{inputenc}
\usepackage{graphicx}
\usepackage[bookmarks=true,bookmarksnumbered=true,bookmarksopen=true,bookmarksopenlevel=1,
 breaklinks=false,pdfborder={0 0 0},pdfborderstyle={},backref=false,colorlinks=false]
 {hyperref}
\hypersetup{pdftitle={The Ethics of ChatGPT in Medicine and Healthcare},
 pdfauthor={Joschka Haltaufderheide and Robert Ranisch},
 pdfpagelayout=OneColumn, pdfnewwindow=true, pdfstartview=XYZ, plainpages=false}

\makeatletter

\pdfpageheight\paperheight
\pdfpagewidth\paperwidth

\let\oldforeign@language\foreign@language
\DeclareRobustCommand{\foreign@language}[1]{%
  \lowercase{\oldforeign@language{#1}}}

\usepackage[caption=false,font=footnotesize]{subfig}
\usepackage{makecell, longtable}

\usepackage{stfloats}

\usepackage{amsmath}
\usepackage[skip=1ex]{caption}
\usepackage{booktabs}
\renewcommand{\appendixname}{Supplement}
\usepackage{doi}

\makeatother

\usepackage[style=nature,doi = true]{biblatex}
\begin{document}
\title{The Ethics of ChatGPT in Medicine and Healthcare: A Systematic Review
on Large Language Models (LLMs) }
\author{Joschka~Haltaufderheide\thanks{Joschka Haltaufderheide is with the Juniorprofessorship for Medical
Ethics with a Focus on Digitization, Faculty of Health Sciences Brandenburg,
University of Potsdam, Germany, e\textendash mail: joschka.haltaufderheide@uni\textendash potsdam.de.
ORCID: 0000\textendash 0002\textendash 5014\textendash 4593.} \& Robert~Ranisch\thanks{Robert Ranisch is Juniorprofessor for Medical Ethics with a Focus
on Digitization, Faculty of Health Sciences Brandenburg, University
of Potsdam, Germany, e\textendash mail: ranisch@uni-potsdam.de. ORCID:
0000\textendash 0002\textendash 1676\textendash 1694} }
\IEEEspecialpapernotice{Preprint Version }
\IEEEaftertitletext{}
\markboth{}{Haltaufderheide \& Ranisch: The Ethics of ChatGPT}
\IEEEpubid{}
\maketitle
\begin{abstract}
Background: With the introduction of ChatGPT, Large Language Models
(LLMs) have received enormous attention in healthcare. Despite their
potential benefits, researchers have underscored various ethical implications.
While individual instances have drawn much attention, the debate lacks
a systematic and comprehensive overview of practical applications
currently researched and ethical issues connected to them. Against
this background, this work aims to map the ethical landscape surrounding
the current stage of deployment of LLMs in medicine and healthcare.

Methods: Electronic databases and commonly used preprint servers were
queried using a comprehensive search strategy which generated 796
records. Studies were screened and extracted following a modified
rapid review approach. Methodological quality was assessed using a
hybrid approach. For 53 records, a meta-aggregative synthesis was
performed.

Results: Four general fields of applications emerged and testify to
a vivid phase of exploration. Advantages of using LLMs are attributed
to their capacity in data analysis, personalized information provisioning,
and support in decision-making or mitigating information loss and
enhancing medical information accessibility. However, our study also
identifies recurrent ethical concerns connected to fairness, bias,
non-maleficence, transparency, and privacy. A distinctive concern
is the tendency to produce harmful misinformation or convincingly
but inaccurate content. A recurrent plea for ethical guidance and
human oversight is evident.

Discussion: Given the variety of use cases, it is suggested that the
ethical guidance debate be reframed to focus on defining what constitutes
acceptable human oversight across the spectrum of applications. This
involves considering the diversity of setting, varying potentials
for harm, and different acceptable thresholds for performance and
certainty in diverse healthcare settings. In addition, a critical
inquiry is necessary to determine the extent to which the current
experimental use of LLMs is both necessary and justified.
\end{abstract}


\IEEEpeerreviewmaketitle{}

\section{Introduction}

\IEEEPARstart{L}{arge} language models (LLMs) have emerged as a
transformative force in artificial intelligence (AI), generating significant
interest across various sectors. The 2022 launch of OpenAI\textquoteright s
ChatGPT demonstrated their groundbreaking capabilities, revealing
the current state of development to a wide audience. Public availability
and scientific interest, since then, have resulted in a flood of scientific
papers considering possible areas of application\supercite{Kaddour_etal_2023}
as well as their ethical and social implications from a practical
perspective.\supercite{Bommasani_etal_2022} A burgeoning field for
LLMs lies within medicine and healthcare,\supercite{Lee_etal_2023f}
encompassing clinical, educational and research applications.\supercite{Lee_etal_2023f,Lee_etal_2023e,Thirunavukarasu_etal_2023,Clusmann_etal_2023,Sallam_2023a,Dave_etal_2023,Korngiebel_Mooney_2021}
In a brief span, a significant number of publications has investigated
the potential uses of LLMs in these domains,\supercite{Li_etal_2023e}
indicating a positive trajectory for the integration of medical AI.
Present-day LLMs, such as ChatGPT, are considered to have a promising
accuracy in clinical decision-making,\supercite{Rao_etal_2023,LIU_etal_2023f}
diagnosis,\supercite{Takita_etal_2024} symptom-assessment, and triage-advise.\supercite{Kim_etal_2024}
In patient-communication, it has been posited that LLMs can also generate
empathetic responses.\supercite{Ayers_etal_2023} LLMs specifically
trained on biomedical corpora forebode even further capacities for
clinical application and patient care\supercite{Singhal_etal_2023}
in the foreseeable future.

Conversely, the adoption of LLMs is entwined with ethical and social
concerns.\supercite{Hagendorff_2024} In their seminal work, Bender
et al. anticipated real-world harms that could arise from the deployment
of LLMs.\supercite{Bender_etal_2021a} Scholars have delineated potential
risks across various application domains.\supercite{Weidinger_etal_2022,Ray_2023}
The healthcare and medical field, being particularly sensitive and
heavily regulated, is notably susceptible to ethical dilemmas. This
sector is also underpinned by stringent ethical norms, professional
commitments, and societal role recognition. Despite the potential
benefits of employing advanced AI technology, researchers have underscored
various ethical implications associated with using LLMs in healthcare
and health-related research.\supercite{Lee_etal_2023e,Clusmann_etal_2023,Gilbert_etal_2023,Li_etal_2023b,Sallam_2023a,Wang_etal_2023e,Grote_Berens_forthcoming}
Paramount concerns include the propensity of LLMs to disseminate inadequate
information, input of sensitive health information or patient data,
raising concerns regarding privacy,\supercite{Li_etal_2023b} and
perpetuating harmful gender, cultural or racial biases,\supercite{Abid_etal_2021,Yeung_etal_2023,Omiye_etal_2023,Zack_etal_2024}
well known from machine learning algorithms\supercite{Suresh_Guttag_2021}
especially in healthcare.\supercite{Obermeyer_etal_2019} Case reports
have documented that ChatGPT has already led to actual real damages,
potentially life-threatening for patients.\supercite{Saenger_etal_2024}

While individual instances have drawn attention to ethical concerns
surrounding the use of LLMs in healthcare, there appears to be a deficit
in comprehensive, systematic overviews addressing these ethical considerations.
This gap is significant, given the ambitions to rapidly integrate
LLMs and foundational models into healthcare systems.\supercite{Moor_etal_2023}
Our intention is to bridge this lacuna by mapping out the ethical
landscape that surrounds the deployment of LLMs in this field. To
achieve this, we will provide an encompassing overview of applications
and ethical considerations pertinent to the utilization of LLMs in
medicine and healthcare. We will, first, provide an encompassing overview
of applications and ethical considerations pertinent to the utilization
of LLMs in medicine and healthcare by delineating the ethically relevant
applications, interventions, and contexts where LLMs have been tested
or proposed within the field. Secondly, we aim to identify the principal
outcomes as well as the opportunities, risks, benefits, and potential
harms associated with the use of LLMs, as deemed significant from
an ethical standpoint. With this, we aspire to not only to outline
the current ethical discourse but also to inform future dialogues
and policy-making in the intersection of LLMs and healthcare ethics.

\section{Methods}

A review protocol focusing on practical applications and ethical considerations
grounded in experience was designed by the authors and registered
in the international prospective register of systematic reviews.\supercite{Ranisch_Haltaufderheide_}
Relevant publication databases and preprint servers were queried.
Inclusions were screened and extracted in a two-staged process following
a modified rapid review approach.\supercite{Garritty_etal_2024} Inclusion
and exclusion criteria were based on the three key concepts of intervention,
application setting, and outcomes (see Supplement 1). No additional
inclusion or exclusion criteria (e.g. publication type) were applied.
However, we excluded work that was solely concerned with (ethical)
questions of medical education, academic writing, authorship and plagiarism.
While we recognize that these issues are affected by the use of LLMs
in significant ways\supercite{Clusmann_etal_2023,Abd-alrazaq_etal_2023,Liebrenz_etal_2023}
these, challenges are not specific to health-related applications.
Database searches were conducted in July 2023 (see Table~\ref{tab:source}
for sources). Subsequently, the authors independently screened titles
and abstracts of 10\% of all database hits (73 records) to test and
refine inclusion and exclusion criteria. After a joint discussion
of the results, the remaining 90\% were screened by the first author.

\begin{table}[ht] 
\caption{\label{tab:source}Overview on Sources and Searchstring}
\begin{tabular}{
>{\raggedright\arraybackslash}p{\dimexpr0.40\linewidth-2\tabcolsep-1.33\arrayrulewidth}
>{\raggedright\arraybackslash}p{\dimexpr0.60\linewidth-2\tabcolsep-1.33\arrayrulewidth}
}

\toprule
\multicolumn{2}{c}{\thead{Sources}}\\ 
\cmidrule(lr){1-2}
Databases & MEDLINE via PubMed\\
 & CINAHL \\
 & Embase \\
 & Philosophers' Index \\
 & PsychInfo \\
 & IEEEX Xplore\\
\cmidrule(lr){1-2}
Preprint Servers & arXiv \\
 & MedRxiv \\
 & BioRxiv \\
 & TechRxiv \\
 & OSF Preprints \\
\toprule
\multicolumn{2}{c}{\thead{Search}}\\ 
\cmidrule(lr){1-2}
 Searchstring & 1. ChatGPT [Text Word] \\
 & 2. LLM [Text Word] \\
 & 3. Large Language Model [Text Word] \\
 & 4. 1 OR 2 OR 3 \\
 & 4. Ethics [Text Word] \\
 & 5. Moral [Text Word] \\
 & 6. 4 OR 5 \\
 & 7. 3 AND 6 \\
\multicolumn{2}{p{\dimexpr\linewidth-2\tabcolsep-1.33\arrayrulewidth}}{%
Wildcards and database-specific truncations (e.g. ethic*, moral*) where used where appropriate and applicable.}\\ 
\bottomrule
\end{tabular}
\end{table}

Data was extracted using a self-designed extraction form (see Supplement
2). The extraction categories were transformed into a coding tree
using MaxQDA. Both authors independently coded 10\% of the material
to develop and refine the coding scheme in more detail. The remaining
material was extracted by J.H. Results were iteratively discussed
in three joint coding sessions.

A final synthesis was conducted following a meta-aggregative approach.
Based on our extraction fields, we, first, developed preliminary categories
encompassing actors, values, device properties, arguments, recommendations
and conclusions. These categories were, then, iteratively refined
and aggregated through additional coding until saturation was reached.

Given the constraints of normative quality appraisal\supercite{Mertz_2019}
and in line with our research goal to portrait the landscape of ethical
discussions, we decided to take a hybrid approach to the quality question.
We descriptively report on procedural quality criteria (see Table~\ref{tab:records})
to distinguish material that underwent processual quality control
(such as peer review) from other material. In addition, we critically
engage with the findings during reporting to appraise comprehensiveness
and validity of the extracted information pieces.

\section{Results}

Our search yielded a total of 796 database hits. After removal of
duplicates, 738 records went through title/abstract screening. 158
full-texts were assessed. 53 records were included in the dataset,
encompassing 23 original articles, \supercite{Wang_etal_2023e,Agbavor_Liang_2022,Ali_etal_2023,Almazyad_etal_2023,Antaki_etal_2023,Connor_ONeill_2023a,Carullo_etal_2023,Ferrara_2023a,Guo_etal_2023c,Harskamp_Clercq_2023,Hosseini_etal_2023,Knebel_etal_2023,DeAngelis_etal_2023,Padovan_etal_2023,Pal_etal_2023,Rau_etal_2023,Schmalzle_Wilcox_2022,Shahriar_Hayawi_2023,Stewart_etal_2022,Suresh_etal_2023,Tang_etal_2023,Yeo_etal_2023a,Yeo_etal_2023}
including theoretical or empirical work, 11 letters,\supercite{Ahn_2023,Arslan_2023,Beltrami_Grant-Kels_2023,Buzzaccarini_etal_2023,Cheng_etal_2023a,Gupta_etal_2023,Howard_etal_2023,Li_etal_2023g,Perlis_2023,Waisberg_etal_2023,Zhong_etal_2023}
six editorials,\supercite{Jairoun_etal_2023,Kavian_etal_2023,Page_etal_2023,Singh_2023,Thomas_2023,Yoder-Wise_2023}
four reviews,\supercite{Sallam_2023,Dave_etal_2023,Temsah_etal_2023,Xie_Wang_2023}
three comments,\supercite{Li_etal_2023b,Abdulai_Hung_2023,Ferreira_Lipoff_2023}
one report\supercite{Guo_etal_2023d} and five unspecified articles.\supercite{Currie_2023,Eggmann_Blatz_2023,Gottlieb.2023,Harrer_2023,Snoswell_etal_2023}
Most works focus on applications utilizing ChatGPT across various
healthcare fields, as indicated in Table~\ref{tab:records}. \begin{table*}[hb] 
\caption{\label{tab:records}Overview on Included Records}
\begin{tabular}{
>{\raggedright\arraybackslash}p{\dimexpr0.20\linewidth-2\tabcolsep-1.33\arrayrulewidth}
>{\raggedright\arraybackslash}p{\dimexpr0.15\linewidth-2\tabcolsep-1.33\arrayrulewidth}
>{\raggedright\arraybackslash}p{\dimexpr0.15\linewidth-2\tabcolsep-1.33\arrayrulewidth}
>{\raggedright\arraybackslash}p{\dimexpr0.12\linewidth-2\tabcolsep-1.33\arrayrulewidth}
>{\raggedright\arraybackslash}p{\dimexpr0.18\linewidth-2\tabcolsep-1.33\arrayrulewidth}
>{\raggedright\arraybackslash}p{\dimexpr0.20\linewidth-2\tabcolsep-1.33\arrayrulewidth}}
\toprule
 \multicolumn{2}{c}{\thead{Publication}} & \multicolumn{2}{c}{\thead{Procedural Quality Control}} & \multicolumn{2}{c}{\thead{Setting}}\\ 
\cmidrule(lr){1-2}\cmidrule(lr){3-4}\cmidrule(lr){5-6}
\thead{\text{Title}}&\thead{Type of Work}  &\thead{Peer Reviewed} &\thead{COI}  &\thead{Device}  &\thead{Field of Application }
\\  \midrule
\citeauthor{Abdulai_Hung_2023}\supercite{Abdulai_Hung_2023} & Commentary & Unclear & Unclear & ChatGPT; ChatGPT 4 & Nursing education, research and practice
\\  \midrule
\citeauthor{Agbavor_Liang_2022}\supercite{Agbavor_Liang_2022} & Empirical Article & Yes & None disclosed & GPT 3 & Neurology 
\\  \midrule
\citeauthor{Ahn_2023}\supercite{Ahn_2023} & Letter & No & None disclosed &  ChatGPT & Emergency Medicine
\\  \midrule
\citeauthor{Ali_etal_2023}\supercite{Ali_etal_2023} & Theoretical Article & Preprint & Unclear &  ChatGPT, Google Bard, Meta LLaMA & Healthcare 
\\  \midrule 
\citeauthor{Almazyad_etal_2023}\supercite{Almazyad_etal_2023} & Empirical Article & Yes & Unclear &  ChatGPT 4 & Pediatric Palliative Care
\\  \hline 
\citeauthor{Antaki_etal_2023}\supercite{Antaki_etal_2023} & Empirical Article & Preprint & Unclear &   ChatGPT; GPT 3.5 & Ophtalmology
\\  \midrule
\citeauthor{Arslan_2023}\supercite{Arslan_2023} & Letter & No & None disclosed &  ChatGPT & Obesity Treatment
\\  \midrule
\citeauthor{Beltrami_Grant-Kels_2023}\supercite{Beltrami_Grant-Kels_2023} & Letter & No & Conflict disclosed &  ChatGPT & Dermatology
\\  \midrule
\citeauthor{Buzzaccarini_etal_2023}\supercite{Buzzaccarini_etal_2023} & Letter & No & Conflict disclosed &  ChatGPT & Aesthetic Medicine
\\  \midrule
\citeauthor{Carullo_etal_2023}\supercite{Carullo_etal_2023} & Empirical Article & Yes & None disclosed & ChatGPT  & Epidemiological Research 
\\  \midrule
\citeauthor{Cheng_etal_2023a}\supercite{Cheng_etal_2023a} & Letter & No & None disclosed &  ChatGPT; GPT 3 & Infectiology
\\  \midrule
\citeauthor{Connor_ONeill_2023a}\supercite{Connor_ONeill_2023a} & Theoretical Article & Preprint & Unclear &  ChatGPT; ChatDoctor; Google BARD & Sport Science and Medicine
\\  \midrule
\citeauthor{Currie_2023}\supercite{Currie_2023} & Unspecified & Yes & None disclosed &  ChatGPT; GPT 3.5 & Nuclear Medicine and Radiology
\\  \midrule
\citeauthor{Dave_etal_2023}\supercite{Dave_etal_2023} & Review & Yes & None disclosed &  ChatGPT & Medicine
\\  \midrule
\citeauthor{DeAngelis_etal_2023}\supercite{DeAngelis_etal_2023} & Theoretical Article & Yes & Conflict disclosed &  GPT; BERT; GPT 2; GPT 3; GPT 4; Instruct GPT; BioBERT; BioGPT; PubMedGPT; Med-PaLm; CORD-19 & Public Health 
\\  \midrule
\citeauthor{Eggmann_Blatz_2023}\supercite{Eggmann_Blatz_2023} & Unspecified & Unclear & None disclosed &  ChatGPT & Dentistry
\\  \midrule
\citeauthor{Ferrara_2023a}\supercite{Ferrara_2023a} & Theoretical Article & Preprint & Unclear &  ChatGPT & Healthcare 
\\  \midrule
\citeauthor{Ferreira_Lipoff_2023}\supercite{Ferreira_Lipoff_2023} & Commentary & Unclear & None disclosed &  ChatGPT & Dermatology
\\  \bottomrule

\multicolumn{6}{r}{\footnotesize\textit{continued on the next page}}
\end{tabular}
\end{table*}

\addtocounter{table}{-1}
\begin{table*}[ht] 
\caption{Overview on Included Records (Continued)}
\begin{tabular}{
>{\raggedright\arraybackslash}p{\dimexpr0.20\linewidth-2\tabcolsep-1.33\arrayrulewidth}
>{\raggedright\arraybackslash}p{\dimexpr0.15\linewidth-2\tabcolsep-1.33\arrayrulewidth}
>{\raggedright\arraybackslash}p{\dimexpr0.15\linewidth-2\tabcolsep-1.33\arrayrulewidth}
>{\raggedright\arraybackslash}p{\dimexpr0.12\linewidth-2\tabcolsep-1.33\arrayrulewidth}
>{\raggedright\arraybackslash}p{\dimexpr0.18\linewidth-2\tabcolsep-1.33\arrayrulewidth}
>{\raggedright\arraybackslash}p{\dimexpr0.20\linewidth-2\tabcolsep-1.33\arrayrulewidth}}
\toprule
 \multicolumn{2}{c}{\thead{Publication}} & \multicolumn{2}{c}{\thead{Procedural Quality Control}} & \multicolumn{2}{c}{\thead{Setting}}\\ 
\cmidrule(lr){1-2}\cmidrule(lr){3-4}\cmidrule(lr){5-6}
\thead{\text{Title}}&\thead{Type of Work}  &\thead{Peer Reviewed} &\thead{COI}  &\thead{Device}  &\thead{Field of Application }
\\  \midrule
\citeauthor{Gottlieb.2023}\supercite{Gottlieb.2023} & Unspecified & Yes & None disclosed & ChatGPT & Emergency Medicine
\\  \midrule
\citeauthor{Guo_etal_2023c}\supercite{Guo_etal_2023c} & Empirical Article & Preprint & Conflict disclosed & ChatGPT; GPT 3; NeuroGPT-X & Neurosurgery
\\  \midrule
\citeauthor{Guo_etal_2023d}\supercite{Guo_etal_2023d} & Report & Preprint & Unclear & ProteinChat & Protein Research
\\  \midrule
\citeauthor{Gupta_etal_2023}\supercite{Gupta_etal_2023} & Letter & No & None disclosed & ChatGPT & Aesthetic Surgery
\\  \midrule
\citeauthor{Harrer_2023}\supercite{Harrer_2023} & Unspecified & Yes & Conflict disclosed & ChatGPT; LaMDA; BARD; Med-Palm & Healthcare
\\  \midrule 
\citeauthor{Harskamp_Clercq_2023}\supercite{Harskamp_Clercq_2023} & Empirical Article & Preprint & None disclosed & ChatGPT; InstructGPT & Cardiopulmonary Medicine
\\  \midrule
\citeauthor{Hosseini_etal_2023}\supercite{Hosseini_etal_2023} & Empirical Article & Preprint & Unclear & ChatGPT; GPT 4; Elicit; Med-PaLM & Education, Research and Healthcare 
\\  \midrule 
\citeauthor{Howard_etal_2023}\supercite{Howard_etal_2023} & Letter & No & Conflict disclosed & ChatGPT & Infection Medicine 
\\  \midrule
\citeauthor{Jairoun_etal_2023}\supercite{Jairoun_etal_2023} & Editorial & No & Unclear & ChatGPT & Pharmacy
\\  \midrule
\citeauthor{Kavian_etal_2023}\supercite{Kavian_etal_2023} & Editorial & No & None disclosed & ChatGPT & Surgery 
\\  \midrule
\citeauthor{Knebel_etal_2023}\supercite{Knebel_etal_2023} & Empirical Article & Preprint & None disclosed & ChatGPT; GPT 3 & Ophtalmology
\\  \midrule
\citeauthor{Li_etal_2023g}\supercite{Li_etal_2023g} & Letter & No & No & ChatGPT & Surgery 
\\  \midrule
\citeauthor{Li_etal_2023b}\supercite{Li_etal_2023b} & Commentary & Unclear & No & ChatGPT; BioGPT; LaMDA; Sparrow; Pangu Alpha; OPT-IML; Megataron Turing MLG & Medicine and Medical Research 
\\  \midrule 
\citeauthor{Padovan_etal_2023}\supercite{Padovan_etal_2023} & Empirical Article & Preprint & None disclosed & ChatGPT & Occupational Medicine
\\  \midrule
\citeauthor{Page_etal_2023}\supercite{Page_etal_2023} & Editorial & No & Conflict disclosed & ChatGPT 4 & Microbial genomics research
\\  \midrule
\citeauthor{Pal_etal_2023}\supercite{Pal_etal_2023} & Empirical Article & Preprint & Unclear & BERT; BioBERT; BioClinicalBERT; SciBERT; UMLS-BERT & Medicine 
\\  \midrule
\citeauthor{Perlis_2023}\supercite{Perlis_2023} & Letter & Preprint & Conflict disclosed & ChatGPT 4 & Psychopharmacology
\\  \midrule
\citeauthor{Rau_etal_2023}\supercite{Rau_etal_2023} & Empirical Article & Preprint & Unclear & ChatGPT; GPT 3.5 Turbo; accGPT & Radiology
\\  \bottomrule 
\multicolumn{6}{r}{\footnotesize\textit{continued on the next page}}
\end{tabular}
\end{table*}

\addtocounter{table}{-1}
\begin{table*}[ht] 
\caption{Overview on Included Records (Continued)}
\begin{tabular}{
>{\raggedright\arraybackslash}p{\dimexpr0.20\linewidth-2\tabcolsep-1.33\arrayrulewidth}
>{\raggedright\arraybackslash}p{\dimexpr0.15\linewidth-2\tabcolsep-1.33\arrayrulewidth}
>{\raggedright\arraybackslash}p{\dimexpr0.15\linewidth-2\tabcolsep-1.33\arrayrulewidth}
>{\raggedright\arraybackslash}p{\dimexpr0.12\linewidth-2\tabcolsep-1.33\arrayrulewidth}
>{\raggedright\arraybackslash}p{\dimexpr0.18\linewidth-2\tabcolsep-1.33\arrayrulewidth}
>{\raggedright\arraybackslash}p{\dimexpr0.20\linewidth-2\tabcolsep-1.33\arrayrulewidth}}
\toprule
 \multicolumn{2}{c}{\thead{Publication}} & \multicolumn{2}{c}{\thead{Procedural Quality Control}} & \multicolumn{2}{c}{\thead{Setting}}\\ 
\cmidrule(lr){1-2}\cmidrule(lr){3-4}\cmidrule(lr){5-6}
\thead{\text{Title}}&\thead{Type of Work}  &\thead{Peer Reviewed} &\thead{COI}  &\thead{Device}  &\thead{Field of Application }
\\  \midrule
\citeauthor{Sallam_2023}\supercite{Sallam_2023} & Review & Preprint & None disclosed & ChatGPT & Healthcare
\\  \midrule
\citeauthor{Schmalzle_Wilcox_2022}\supercite{Schmalzle_Wilcox_2022} & Theoretical Article & Yes & None disclosed & GPT 2 & Public Health
\\  \midrule
\citeauthor{Shahriar_Hayawi_2023}\supercite{Shahriar_Hayawi_2023} & Theoretical Article & Preprint & None disclosed & ChatGPT; BERT & Healthcare
\\  \midrule 
\citeauthor{Singh_2023}\supercite{Singh_2023} & Editorial & No & Unclear & ChatGPT & Mental Health
\\  \midrule
\citeauthor{Snoswell_etal_2023}\supercite{Snoswell_etal_2023} & Unspecified & No & Unclear & ChatGPT & Pharmacy
\\  \midrule 
\citeauthor{Stewart_etal_2022}\supercite{Stewart_etal_2022} & Theoretical Article & Preprint & None disclosed & BERT; various Natural language processing models & Healthcare
\\  \midrule
\citeauthor{Suresh_etal_2023}\supercite{Suresh_etal_2023} & Empirical Article & Preprint & None disclosed & ChatGPT; GPT 4 & Otolaryngology 
\\  \midrule 
\citeauthor{Tang_etal_2023}\supercite{Tang_etal_2023} & Empirical Article & Preprint & Unclear & ChatGPT; GPT 3.5 & Medicine
\\  \midrule
\citeauthor{Temsah_etal_2023}\supercite{Temsah_etal_2023} & Review & Yes & None disclosed & ChatGPT & Healthcare and Health Research
\\  \hline 
\citeauthor{Thomas_2023}\supercite{Thomas_2023} & Editorial & No & Unclear & ChatGPT & Mental Health Nursing
\\  \midrule
\citeauthor{Waisberg_etal_2023}\supercite{Waisberg_etal_2023} & Letter & No & None disclosed & ChatGPT; GPT 4 & Opthalmology 
\\  \midrule
\citeauthor{Xie_Wang_2023}\supercite{Xie_Wang_2023} & Review & Preprint & None disclosed &  BERT; BioBERT; BlueBERT; PubMedBERT; ChatGPT; GPT 4; BioGPT; Med-PaLM & Healthcare and Medicine 
\\  \midrule 
\citeauthor{Yeo_etal_2023}\supercite{Yeo_etal_2023} & Empirical Article & Preprint & None disclosed & ChatGPT; GPT 3.5 & Hepatology 
\\  \midrule
\citeauthor{Yeo_etal_2023a}\supercite{Yeo_etal_2023a} & Empirical Article & Preprint & None disclosed & ChatGPT; GPT 4 & Hepatology 
\\  \midrule
\citeauthor{Yeung_etal_2023}\supercite{Yeung_etal_2023} & Empirical Article & Preprint & Unclear & ChatGPT; Foresight; PaLM, Gopher; Chinchilla & Medicine
\\  \midrule 
\citeauthor{Yoder-Wise_2023}\supercite{Yoder-Wise_2023} & Editorial & No & None disclosed & ChatGPT & Nursing
\\  \midrule
\citeauthor{Zhong_etal_2023}\supercite{Zhong_etal_2023} & Letter & No & None disclosed & ChatGPT & Neuropsychiatric practice and research
\\  \bottomrule
\end{tabular}
\end{table*}

During analysis, four general themes emerged in our dataset, which
we use to structure reporting. These themes include clinical applications,
patient support applications, support of health professionals, and
public health perspectives. Table~\ref{tab:exemplar} provides exemplary
scenarios for each theme derived from the dataset.

\begin{table}[hb] 
\caption{\label{tab:exemplar}Exemplary applications of LLMs}

\begin{tabular}{
>{\raggedright\arraybackslash}p{\dimexpr0.18\linewidth-2\tabcolsep-1.33\arrayrulewidth}
>{\raggedright\arraybackslash}p{\dimexpr0.82\linewidth-2\tabcolsep-1.33\arrayrulewidth}
}
\toprule
\multicolumn{2}{c}{\thead{Predictive Analysis and Risk Assessment}}\\ 
\cmidrule(lr){1-2}
\citeauthor{Connor_ONeill_2023a}\supercite{Connor_ONeill_2023a} & Supporting initial diagnose and triaging of patients by fine-tuning LLMs on a specialised dataset of electronic medical records, clinical notes, Sports science and medicine literature.\\
\cmidrule(lr){1-2}
\citeauthor{Stewart_etal_2022}\supercite{Stewart_etal_2022} & Using traditional and modern natural language processing to triage patients on arrival based on structured data and unstructered free-text history of presenting complaint to predict risk stratification. This includes predictions on the likelihood of admission to hospital, prediction of critical illness, prediction of triage score, prediction of provider-assigned chief complaint, prediction of investigation, and prediction of infection.\\
\midrule
\multicolumn{2}{c}{\thead{Patient Consultation and Communication}}\\ 
\cmidrule(lr){1-2}
\citeauthor{Buzzaccarini_etal_2023}\supercite{Buzzaccarini_etal_2023} & Enhancing patient consultations by providing accurate and reliable information on aesthetic procedures, their risks, benefits and potential outcomes, enabling well-informed decisions and improved treatment outcomes.\\
\cmidrule(lr){1-2}
\citeauthor{Currie_2023}\supercite{Currie_2023} & Providing language translation and helping health professionals to communicate with patients speaking foreign languages; helping health professionals to educate their patients and empower patients to take an active role.\\

\bottomrule
\end{tabular}
\end{table}\addtocounter{table}{-1}
\begin{table}[ht]
\caption{Exemplary applications of LLMs (Continued)}
\begin{tabular}{
>{\raggedright\arraybackslash}p{\dimexpr0.18\linewidth-2\tabcolsep-1.33\arrayrulewidth}
>{\raggedright\arraybackslash}p{\dimexpr0.82\linewidth-2\tabcolsep-1.33\arrayrulewidth}
}

\toprule
\multicolumn{2}{c}{\thead{Diagnosis}}\\ 
\cmidrule(lr){1-2}
\citeauthor{Agbavor_Liang_2022}\supercite{Agbavor_Liang_2022} & Using GPT 3 to distinguish individuals with Alzheimers Disease from healthy controls and to infer cognitive testing scores based on linguistic features. It is shown that the approach outperforms conventional approaches and performs comparable to specifically fine-tuned models. Usable as a web app in the doctors office.\\  \midrule
\citeauthor{Rau_etal_2023}\supercite{Rau_etal_2023}& Supporting radiologists diagnostic performance by providing imaging recommendations in accordance with recent guidelines.\\
\toprule
\multicolumn{2}{c}{\thead{Treatment Planning}}\\ 
\cmidrule(lr){1-2}
\citeauthor{Arslan_2023}\supercite{Arslan_2023} & Using ChatGPT to provide personalized recommendations on topics such as nutrition, exercise and psychological support in obesity treatment.\\
\midrule
\citeauthor{Cheng_etal_2023a}\supercite{Cheng_etal_2023a} & Using ChatGPT to provide treatment recommendations based on patients clinical presentation, disease severity, and comorbidities.\\
\toprule
\multicolumn{2}{c}{\thead{Patient Support}}\\ 
\cmidrule(lr){1-2}
\citeauthor{Yeo_etal_2023a}\supercite{Yeo_etal_2023a} & Using ChatGPT as an informational plattform to comprehend and to respond to cirrhosis related questions in different languages, adressing barriers that may impact patient care.\\
\midrule
\citeauthor{Knebel_etal_2023}\supercite{Knebel_etal_2023} & Using ChatGPT for the assessment of acute ophtalmological conditions with regard to triage accurracy and recommendations of preclinical measures.\\
\toprule
\multicolumn{2}{c}{\thead{Professional Support and Reserach}}\\ 
\cmidrule(lr){1-2}
\citeauthor{Hosseini_etal_2023}\supercite{Hosseini_etal_2023} & Using LLMs to increase efficiency in note-taking through prepopulation of forms, voice recording and morphing into clinical notes or synthesizing existing patient notes to save clinicians time.\\
\midrule
\citeauthor{Gottlieb.2023}\supercite{Gottlieb.2023} & Using Conversational AI to create study documents by translating complex concepts into simpler ones or designing informed consent documents for patients.\\
\midrule
\citeauthor{Guo_etal_2023d}\supercite{Guo_etal_2023d} & Using a ChatGPT-like (ProteinGPT) systems to accelerate protein research. The model is aimed at learning and understanding protein 3D structures. ProteinGPT enables users to upload proteins, ask questions, and engage in interactive conversations to gain insights.\\
\toprule
\multicolumn{2}{c}{\thead{Public Health}}\\ 
\cmidrule(lr){1-2}
\citeauthor{Schmalzle_Wilcox_2022}\supercite{Schmalzle_Wilcox_2022}& Using LLMs to create an AI-guided message creation system to disseminate health related information via social media.\\
\midrule
\citeauthor{Cheng_etal_2023a}\supercite{Cheng_etal_2023a} & Using ChatGPT to monitor news and social media platforms for signs of outbreaks of disease clusters and to alert health professionals to potential threats.\\
\bottomrule
\end{tabular}
\end{table}

\subsection{Clinical applications}

\subsubsection{Predictive analysis and risk assessment}

To support initial diagnose and triaging of patients,\supercite{Connor_ONeill_2023a,Stewart_etal_2022}
several authors discuss the use of LLMs in the context of predictive
patient analysis and risk assessment in or prior to clinical situations
as a potentially transformative application.\supercite{Sallam_2023,Currie_2023}
The role of LLMs in this scenario is described as that of a \textquotedblleft co-pilot\textquotedblright{}
using available patient information to flag areas of concern or to
predict diseases and risk factors.\supercite{Hosseini_etal_2023}

\citeauthor{Currie_2023}, in line with most authors, notes that predicting
health outcomes and relevant patterns is very likely to improve patient
outcomes and contributes to patient benefit.\supercite{Currie_2023}
For example, overcrowded emergency departments present a serious issue
worldwide and have a significant impact on patient outcomes. From
a perspective of avoidance of harm, using LLMs with triage notes could
lead to reduced length of stay and a more efficient utilization of
time in the waiting room.\supercite{Stewart_etal_2022}

All authors note, however, that such applications might also be problematic
and require close human oversight.\supercite{Connor_ONeill_2023a,Hosseini_etal_2023,Shahriar_Hayawi_2023,Currie_2023}
Although LLMs might be able to reveal connections between disparate
knowledge,\supercite{Carullo_etal_2023} generating inaccurate information
would have severe negative consequences.\supercite{Hosseini_etal_2023,Sallam_2023}
It might lead to direct harm to patients or provide clinicians with
false and dangerous justifications and rationales for their decisions.\supercite{Sallam_2023}
These problems are tightly connected to inherent biases in LLMs, their
tendency to \textquotedblleft hallucinate\textquotedblright{} and
their intransparency.\supercite{Stewart_etal_2022} In addition, uncertainties
are increased by use of unstructured data. Medical notes often differ
from the data pretrained models utilise. This makes it difficult to
predict accuracy of output when inputting such data or using it to
fine-tune LLMs.\supercite{Stewart_etal_2022} Interpretability of
results and recommendations introduce additional complexity and sources
of potential harm.\supercite{Stewart_etal_2022} \citeauthor{Currie_2023}
notes that despite such difficulties, the use of LLMs proceeds largely
in absence of guidelines, recommendations and control. The outcome,
hence, ultimately depends on clinicians\textquoteright{} ability to
interpret findings and identify inaccurate information.\supercite{Currie_2023}

\subsubsection{Patient consultation and communication}

LLMs can offer a novel approach in patient-provider interaction, where
they can facilitate informational exchange and bridge gaps between
clinical and preclinical settings such as self-management measures
or community aids.\supercite{Dave_etal_2023} This includes easing
the transition between settings by removing barriers to communication\supercite{Hosseini_etal_2023,Buzzaccarini_etal_2023,Currie_2023,Harrer_2023}
or removing barriers in the clinical workflow to facilitate timely
and efficient support. As is suggested, LLMs can collect information
from patients or provide additional information, enabling well-informed
decisions and increasing satisfaction in patients.\supercite{Yeo_etal_2023,Buzzaccarini_etal_2023,Currie_2023}
Provision of language translation and simplification of medical jargon
may allow patients to become more engaged in the process and enhance
patient-provider communication.\supercite{Currie_2023,Harrer_2023}
However, it remains unclear in our dataset how such applications would
look like in practice \textemdash{} specifically where, when and how
LLMs actually could be integrated.

These suggestions require to consider ethically relevant boundaries
regarding the protection of patient data, and safety,\supercite{Ali_etal_2023,Buzzaccarini_etal_2023,Abdulai_Hung_2023,Harrer_2023}
potentially unjust disparities,\supercite{Ali_etal_2023,Buzzaccarini_etal_2023,Harrer_2023}
and the wider dimensions of care such as the therapeutic relationship.\supercite{Ali_etal_2023,Beltrami_Grant-Kels_2023,Cheng_etal_2023a,Li_etal_2023g,Abdulai_Hung_2023}
Robust measures to avoid incorrect information in technological mediation
of communication and the need to strike a balance with \textquotedblleft the
human touch\textquotedblright{} of care\supercite{Buzzaccarini_etal_2023}
are stressed. With regard to the former, \citeauthor{Buzzaccarini_etal_2023}
argue for robust expert oversight. Regarding the latter, \citeauthor{Li_etal_2023g}
note a potential shift in power dynamics between patients and providers
in which providers might lose their authoritative position and might
be seen as less knowledgeable.\supercite{Li_etal_2023g} Others fear
a loss of personal care that should be avoided\supercite{Ali_etal_2023,Cheng_etal_2023a,Abdulai_Hung_2023}
and the lack of contextual content of individual health challenges.\supercite{Abdulai_Hung_2023,Guo_etal_2023c}
Open communication and consent to technical mediation of patient-provider
communication is required to promote trust but might be difficult
to achieve.\supercite{Kavian_etal_2023,Ferreira_Lipoff_2023}

\subsubsection{Diagnosis}

Many studies in our dataset discusses the possible use of LLMs for
diagnosis.\supercite{Dave_etal_2023,Ali_etal_2023,Connor_ONeill_2023a,Hosseini_etal_2023,Beltrami_Grant-Kels_2023,Cheng_etal_2023a,Waisberg_etal_2023,Zhong_etal_2023,Sallam_2023,Temsah_etal_2023,Ferreira_Lipoff_2023,Currie_2023}
It is suggested that the LLMs\textquoteright{} ability to analyze
large amounts of unstructured data provides pathways to timely, efficient
and more accurate diagnosis to the benefit of patients.\supercite{Agbavor_Liang_2022,Ali_etal_2023,Zhong_etal_2023,Temsah_etal_2023,Ferreira_Lipoff_2023}
It might also enable the discovery of hidden patterns\supercite{Connor_ONeill_2023a}
and reduce healthcare costs.\supercite{Ali_etal_2023,Rau_etal_2023}

An ethical problem emerges with potentially negative effects on patient
outcomes due to biases in the training data,\supercite{Ali_etal_2023,Connor_ONeill_2023a,Ferrara_2023a,Sallam_2023,Temsah_etal_2023,Ferreira_Lipoff_2023}
especially with the lack of diverse datasets risking underrepresentation
of marginalized or vulnerable groups. Biased models may result in
unfair treatment of disadvantaged groups, leading to disparities in
access, exacerbating existing inequalities, or harming persons through
selective accuracy.\supercite{Ferrara_2023a} Based on an experimental
study setup, \citeauthor{Yeung_etal_2023} deliver an insightful example
showing that ChatGPT and Foresight NLP exhibit racial bias towards
black patients.\supercite{Yeung_etal_2023} Problems of interpretability,
hallucinations, and falsehood mimicry pile onto this issue and increase
these risks.\supercite{Agbavor_Liang_2022,Ali_etal_2023,Hosseini_etal_2023,Sallam_2023}
With regard to transparency, two sources suggest that LLM-supported
diagnosis hamper the process of providing adequate justification due
to their opacity.\supercite{Ali_etal_2023,Sallam_2023} This is understood
to threaten the authoritative position of professionals, leaving them
at risk of not being able to provide a rationale for a diagnosis\supercite{Agbavor_Liang_2022}
and might lead to an erosion of trust between both parties. This is
in line with others noting that LLMs are not able to replicate a process
of clinical reasoning in general and, hence, fail to comprehend the
complexity of the process.\supercite{Hosseini_etal_2023,Beltrami_Grant-Kels_2023,Temsah_etal_2023}
Based on the principle of avoidance of harm, it is an important requirement
to subject each generated datum to clinical validation as well as
to develop \textquotedblleft ethical and legal systems\textquotedblright{}
to mitigate these problems.\supercite{Ali_etal_2023,Connor_ONeill_2023a,Beltrami_Grant-Kels_2023}

It needs to be noted, however, that the technically unaided process
of diagnoses is also known to be subjective and prone to error.\supercite{Zhong_etal_2023}
This implies that an ethical evaluation should be carried out in terms
of relative reliability and effectiveness compared to existing alternatives.
Whether and under what circumstances this might be the case is a question
that is not addressed.

\subsubsection{Treatment planning}

Six studies in our dataset highlight the use of LLMs in providing
personalized recommendations for treatment regimens or to support
clinicians in treatment decisions based on electronic patient information
or history,\supercite{Arslan_2023,Buzzaccarini_etal_2023,Cheng_etal_2023a,Waisberg_etal_2023,Zhong_etal_2023,Currie_2023}
providing a quick and reliable course of action to clinicians and
patients. However, as with diagnostic applications, biases and perpetuating
existing stereotypes and disparities is a constantly discussed theme.\supercite{Buzzaccarini_etal_2023,Cheng_etal_2023a,Zhong_etal_2023}
\citeauthor{Ferrara_2023a} also cautions that LLMs will likely prioritize
certain types of treatments or interventions over others, disproportionately
benefiting certain groups and disadvantaging others.\supercite{Ferrara_2023a}

Additionally, it is highlighted that inputting patient data raises
ethical questions regarding confidentiality, privacy, and data security.\supercite{Arslan_2023,Buzzaccarini_etal_2023,Cheng_etal_2023a,Waisberg_etal_2023,Zhong_etal_2023}
This especially applies to commercial and publicly available models
such as ChatGPT. Inaccuracies in potential treatment recommendations
are also noted as a concerning source for harm.\supercite{Arslan_2023,Buzzaccarini_etal_2023,Cheng_etal_2023a,Waisberg_etal_2023,Zhong_etal_2023}
In a broader context, several authors suggest that for some LLMs the
absence of internet access, insufficient domain-specific data, limited
access to treatment guidelines, knowledge on local or regional characteristics
of the healthcare system, and outdated research significantly heighten
the risk of inaccurate recommendations.\supercite{Li_etal_2023b,Almazyad_etal_2023,Antaki_etal_2023,Carullo_etal_2023,Padovan_etal_2023,Yeo_etal_2023a}

\subsection{Patient support applications}

\subsubsection{Patient information and education}

Almost all authors concerned with patient-facing applications highlight
the benefits of rapid and timely information access that users experience
with state-of-the-art LLMs. \citeauthor{Kavian_etal_2023} compare
patients\textquoteright{} use of chatbots with shifts that have accompanied
the development of the internet as patient information source.\supercite{Kavian_etal_2023}
Such access can improve laypersons\textquoteright{} health literacy
by providing a needs-oriented access to comprehensible medical information,\supercite{Jairoun_etal_2023}
which is regarded as an important precondition of autonomy to allow
more independent, health-related decisions.\supercite{Dave_etal_2023,Sallam_2023}
In their work on the use of ChatGPT 4 in overcoming language barriers,
\citeauthor{Yeo_etal_2023} highlight an additional benefit, as LLMs
could provide cross-lingual translation and thus contribute to equalizing
healthcare and racial disparities.\supercite{Yeo_etal_2023}

Regarding ethical concerns and risks, biases are seen as an significant
source of harm.\supercite{Dave_etal_2023,Connor_ONeill_2023a,Sallam_2023,Temsah_etal_2023}
The literature also highlights a crucial difference in the ethical
acceptability of using patient support applications, leading to more
critical stance when LLMs are used by laypersons compared to health
professionals.\supercite{Yeung_etal_2023,Suresh_etal_2023} However,
ethical acceptability varies across fields; for instance, otolaryngology
and infectious disease studies find ChatGPT\textquoteright s responses
to patients lack detail but aren\textquoteright t harmful,\supercite{Suresh_etal_2023}
whereas pharmacology and mental health indicate greater potential
risks.\supercite{Zhong_etal_2023,Jairoun_etal_2023} 

\subsubsection{Symptom assessment and health management}

LLMs can offer laypersons personalized guidance, such as lifestyle
adjustments during illness,\supercite{Currie_2023} self-assessment
of symptoms,\supercite{Cheng_etal_2023a,Howard_etal_2023} self-triaging,
and emergency management steps.\supercite{Dave_etal_2023,Ahn_2023}
Although current arrangements seem to perform well and generate compelling
responses,\supercite{Dave_etal_2023,Padovan_etal_2023,Howard_etal_2023}
a general lack of situational awareness is noted as a common problem
that might lead to severe harm.\supercite{Dave_etal_2023,Cheng_etal_2023a,Howard_etal_2023}
Situational awareness means the ability to generate responses based
on contextual criteria such as the personal situation, medical history
or social situation. The inability of most current LLMs to seek clarifications
by asking questions and their lack of sensitivity to query variations
can lead to imprecise answers.\supercite{Knebel_etal_2023,Howard_etal_2023}
For instance, research by \citeauthor{Knebel_etal_2023} on self-triaging
in ophthalmologic emergencies indicates that ChatGPT\textquoteright s
responses can\textquoteright t reliably prioritize urgency, reducing
their usefulness.\supercite{Knebel_etal_2023}

\subsection{Support of health professionals and researchers}

\subsubsection{Documentation and administrative tasks}

LLMs could automate tasks like medical reporting,\supercite{Currie_2023}
or summarizing patient interactions\supercite{Dave_etal_2023} including
automatic population of forms or discharge summaries. The consensus
is that LLMs could streamline clinical workflows,\supercite{Dave_etal_2023,Ali_etal_2023,Harskamp_Clercq_2023,Shahriar_Hayawi_2023,Stewart_etal_2022,Buzzaccarini_etal_2023,Jairoun_etal_2023,Sallam_2023,Currie_2023,Eggmann_Blatz_2023,Harrer_2023}
offering time savings for health professionals currently burdened
with extensive administrative duties.\supercite{Jairoun_etal_2023,Harrer_2023}
By automating these repetitive tasks, professionals could dedicate
more time to high-quality medical tasks.\supercite{Harrer_2023} Crucially,
such applications would require the large-scale integration of LLMs
into existing clinical data systems.\supercite{Rau_etal_2023}

\subsubsection{Research}

In health research, LLMs are suggested to support text, evidence or
data summarization,\supercite{Tang_etal_2023,Li_etal_2023g,Gottlieb.2023}
identify research targets,\supercite{Dave_etal_2023,Cheng_etal_2023a,Thomas_2023,Harrer_2023}
designing experiments or studies,\supercite{Thomas_2023,Harrer_2023}
or facilitate knowledge sharing between collaborators,\supercite{Almazyad_etal_2023,Page_etal_2023,Currie_2023}
and to communicate results.\supercite{Sallam_2023} This highlights
the potentials for accelerating research\supercite{DeAngelis_etal_2023,Guo_etal_2023d}
and relieving researchers of workload,\supercite{Dave_etal_2023,Carullo_etal_2023,Li_etal_2023g,Sallam_2023,Temsah_etal_2023,Harrer_2023}
leading to more efficient research workflows allowing researchers
to spend less time on burdensome routine work.\supercite{Dave_etal_2023,Currie_2023}
To certain authors, this could involve condensing crucial aspects
of their work, like crafting digestible research documents for ethics
reviews or consent forms.\supercite{Gottlieb.2023} However, LLMs
capacities are also critically examined, with \citeauthor{Tang_etal_2023}
Tang et al. emphasizing ChatGPT\textquoteright s tendency to produce
attribution and misinterpretation errors, potentially distorting original
source information, echoing concerns over interpretability, reproducibility,
uncertainty handling, and transparency.\supercite{Tang_etal_2023,Sallam_2023}

Some authors fear that using LLMs could compromise research integrity
by disrupting traditional trust factors like source traceability,
factual consistency, and process transparency.\supercite{Li_etal_2023b}
Additionally, concerns about overreliance and deskilling are raised,
as LLMs might diminish researchers\textquoteright{} skills and overly
shape research outcomes.\supercite{DeAngelis_etal_2023} Given that
using such technologies inevitably introduces biases and distortions
to the research flow, \citeauthor{Page_etal_2023} suggest researchers
must maintain vigilance to prevent undue influence from biases introduced
by these technologies, advocating for strict human oversight and revalidation
of outputs.\supercite{Page_etal_2023}

\begin{figure*}[tbh]
\includegraphics[width=182mm]{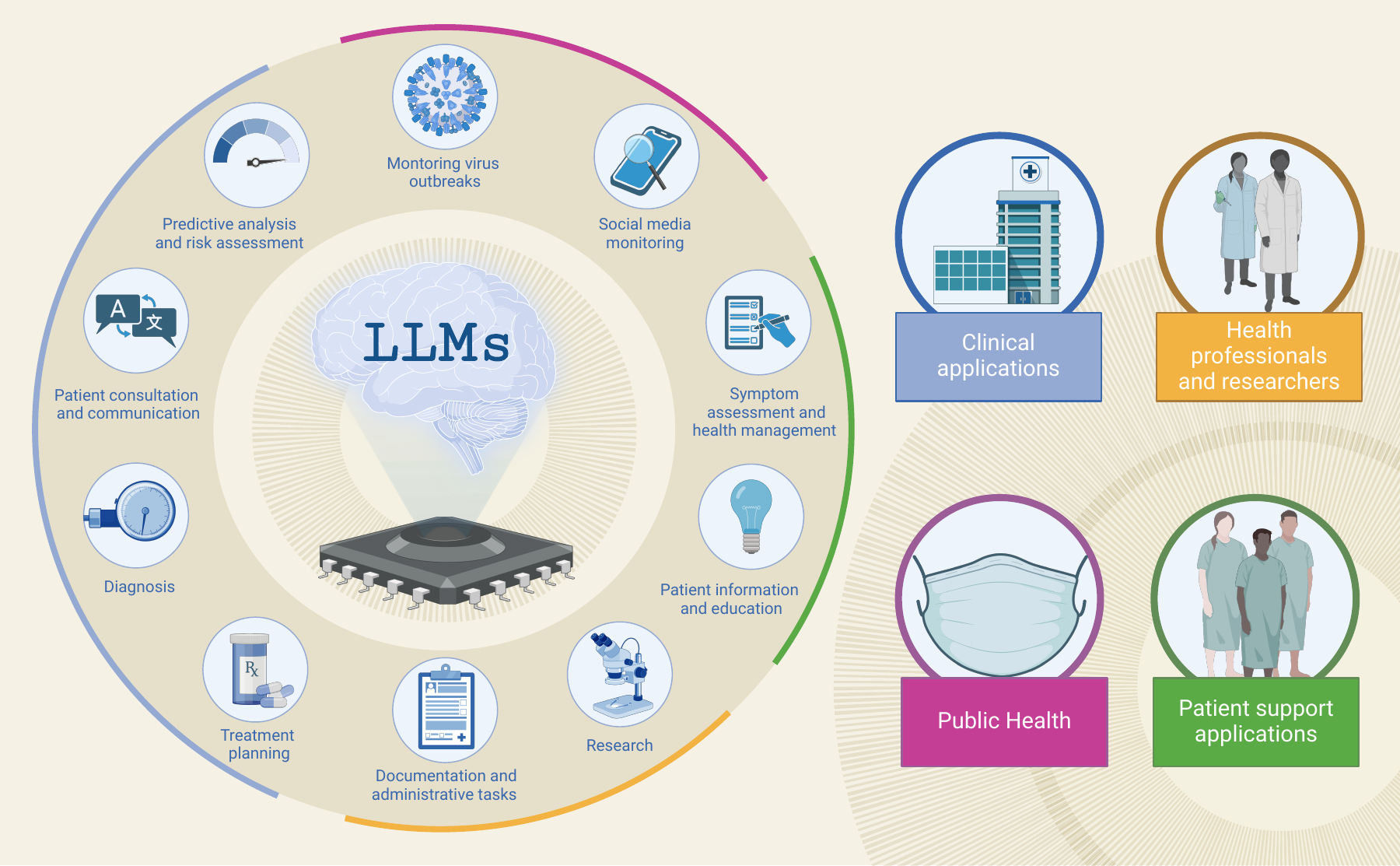}

\caption{Fields of Application of LLMs in Medicine}
\end{figure*}

\subsection{Public Health perspectives}

The dataset encompasses studies that explore the systemic implications
of LLMs, especially from a public health perspective.\supercite{Schmalzle_Wilcox_2022,Cheng_etal_2023a,Temsah_etal_2023}
This includes using LLMs in public health campaigns, for monitoring
news and social media for signs of disease outbreaks\supercite{Cheng_etal_2023a}
and targeted communication strategies.\supercite{Schmalzle_Wilcox_2022}
Additionally, research examines the potential for improving health
literacy or access to health information, especially in low-resource
settings. Access to health information through LLMs can be maintained
free of charge or at very low costs for laypersons.\supercite{Yeo_etal_2023a}
Considering the case of mental health, especially low and middle income
countries might benefit.\supercite{Singh_2023} These countries often
have a huge treatment gap driven by a deficit in professionals or
inequitable resource distribution. Using LLMs could mitigate accessibility
and affordability issues, potentially offering a more favorable alternative
to the current lack of access.\supercite{Singh_2023}

However, a number of authors raise doubts about overly positive expectations.
\citeauthor{Schmalzle_Wilcox_2022} highlight the risks of a dual
use of LLMs.\supercite{Schmalzle_Wilcox_2022} While they might further
equal access to information, malicious actors can and seem to be using
LLMs to spread fake information and devise health messages at an unprecedented
scale that is harmful to societies.\supercite{Schmalzle_Wilcox_2022,Shahriar_Hayawi_2023,Temsah_etal_2023}
\citeauthor{DeAngelis_etal_2023} take this concern one step further,
presenting the concept of an AI-driven infodemic\supercite{DeAngelis_etal_2023}
in which the overwhelming spread of imprecise, unclear, or false information
leads to disorientation and potentially harmful behavior among recipients.
Health authorities have often seen AI technologies as solutions to
information overload. However, the authors caution that an AI-driven
infodemics could exacerbate future health threats. While infodemic
issues in social media and grey literature are noted, AI-driven infodemics
could also inundate scientific journals with low-quality, excessively
produced content.\supercite{DeAngelis_etal_2023}

The commercial nature of most current LLMs systems present another
critical consideration. The profit-driven nature of the field can
lead to concentrations of power among a limited number of companies
and a lack of transparency. This economic model, as highlighted by
several studies, can have negative downstream effects on accessibility
and affordability.\supercite{Li_etal_2023b,Ali_etal_2023,Harskamp_Clercq_2023}
Developing, using, or refining models can be expensive, limiting accessibility
and customization for marginalized communities. Power concentration
also means pricing control lies with LLM companies, with revenues
predominantly directed towards them.\supercite{Hosseini_etal_2023}
These questions are also mirrored in the selection of training data
and knowledge bases\supercite{Li_etal_2023b} which typically encompass
knowledge from well-funded, English speaking countries and, thus,
significantly underrepresents knowledge from other regions. This could
exacerbate health disparities by reinforcing biases rather than alleviating
them.

\section{Discussion}

Our analysis has unveiled an extensive range of LLM applications currently
under investigation in medicine and healthcare (see Figure 1). This
surge in LLMs was largely caused by the advent and ease of use of
ChatGPT, a platform not originally tailored for professional healthcare
settings, yet widely adopted within it.\supercite{Li_etal_2023e,Harrer_2023}
This presents a rather unique instance where a general-purpose technology
has rapidly permeated the sector of healthcare and research to an
unprecedented extent.
\begin{figure*}[tbh]
\includegraphics[width=182mm]{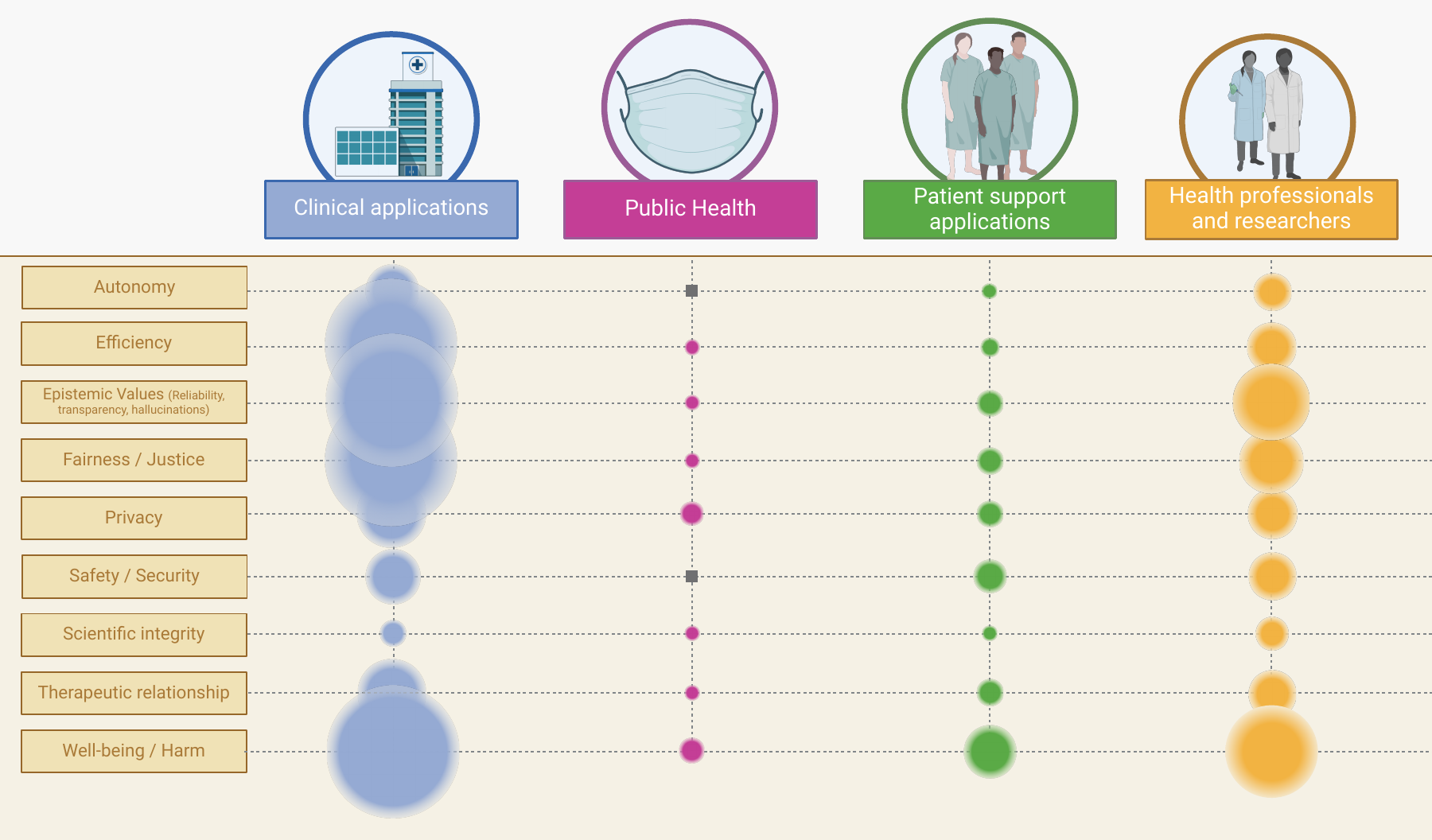}

\caption{Discussed Dimensions of Impact of LLMs}
\end{figure*}

Our review highlights a vivid testing phase of LLMs across various
healthcare domains.\supercite{Li_etal_2023e} Despite the lack of
real-world applications, especially in the clinic, there is an overarching
sentiment of the promise LLMs hold. It is posited that these tools
could increase the efficiency of healthcare delivery and research,
with the potential to benefit patient outcomes while alleviating burdensome
workload of healthcare professionals. These advantages of LLMs are
largely attributed to their capabilities in data analysis, personalized
information provisioning, and support in decision-making, particularly
where quick analysis of voluminous unstructured data is paramount.
Moreover, by mitigating information loss and enhancing medical information
accessibility, LLMs stand to significantly bolster healthcare quality.

However, our study has also surfaced recurrent ethical concerns associated
with LLMs. These concerns echo the wider discourse on AI ethics,\supercite{Kazim_Koshiyama_2021,Hagendorff_2020,Jobin_etal_2019}
particularly in healthcare,\supercite{Morley_etal_2020a} and touch
on issues of fairness, bias, non-maleficence, transparency, and privacy.
Yet, LLMs introduce a distinctive concern linked to a dimension of
epistemic values, that is, their tendency to produce harmful misinformation
or convincingly but inaccurate content through hallucinations as illustrated
in Figure 2.\supercite{Xu_etal_2024} The effects of such misinformation
are particularly severe in healthcare, where the outcome could be
dire. The inherent statistical and predictive architecture combined
with the intransparency of LLMs presents significant hurdles in validating
the clinical accuracy and reliability of their outputs.\supercite{Grote_Berens_2020,Grote_2021,Babushkina_Votsis_2022}

The inclination of LLMs to output erroneous information underscores
the need for human oversight and continual validation of machine-generated
output, as our dataset demonstrates. This need is accentuated by the
lack of professional guidelines or regulatory oversight within this
field.\supercite{Gilbert_etal_2023} Consequently, there is a noticeable
demand for ethical guidelines, evidenced within the literature surrounding
healthcare applications of LLMs.\supercite{DeAngelis_etal_2023,Buzzaccarini_etal_2023,Li_etal_2023g,Page_etal_2023,Singh_2023,Sallam_2023,Temsah_etal_2023,Ferreira_Lipoff_2023}

While we concur with the need for such guidance, our analysis suggests
that the real challenge lies not in the articulation of such a need
but in comprehending the scope of what this entails. There are inherent
and contextual limitations and benefits associated with LLMs that
warrant consideration. Inherently, state-of-the-art LLMs carry the
risks of biases, hallucinations, and challenges in validity assessment,
reliability testing, and reproducibility. Contextually, the effectiveness
of LLM usage hinges on various situational factors, including the
user utilizing LLMs, their level of expertise as well as their epistemic
position (e.g. expert versus layperson), the specific domain of application,
the risk profile of the application, and potential alternatives that
the LLM is compared against.

A nuanced ethical discourse must recognize the multilayered nature
of LLM\textendash usage, from the epistemic stance of the user to
the potential for harm, the varying degrees of potential harm due
to misinformation or bias, and the diverse normative benchmarks for
performance and acceptable levels of uncertainty. Our recommendation
is to reframe the ethical guidance debate to focus on defining what
constitutes acceptable human oversight and validation across the spectrum
of applications and users. This involves considering the diversity
of epistemic positions of users, the varying potentials for harm,
and the different acceptable thresholds for performance and certainty
in diverse healthcare settings.

Given these questions, a critical inquiry is necessary to the extent
to which the current experimental use of LLMs is both necessary and
justified. Our dataset exemplifies a diversity of perspectives, methodologies,
and applications of LLMs, revealing a significant degree of ambiguity
and uncertainty about the appropriate engagement with this technology.
Notably, a portion of current research seems propelled more by a sense
of experimental curiosity than by well-defined methodological rigor,
at times pushing the boundaries of ethical acceptability, particularly
when sensitive real patient data are utilized to explore capabilities
of systems like ChatGPT.

To frame these developments, it is instructive to adopt the perspective
of viewing the implementation of LLMs as akin to a \textquotedblleft social
experiment\textquotedblright .\supercite{vandePoel_2013,vandePoel_2016}
We employ this concept in a descriptive sense to denote a situation
in which \textendash{} according to van der Poel \textendash{} the
benefits, risks and ethical issues of a technology can only fully
manifest subsequent to its widespread introduction.\supercite{vandePoel_2016}
Such a stance recognizes that the novelty of LLMs, combined with its
inherent complexity and opacity, necessitates an iterative process
of diminishing uncertainties and learning through which consequences
only gradually emerge. By the same time, framing the current developments
as social experiment also reinforces the need to establish and respect
ethical limits \textendash{} especially within the healthcare domain,
where professional duties and responsibilities towards patients are
foundational.

With this in mind we suggest that understanding how we acquaint ourselves
with disruptive technologies must be central to any future ethical
discourse. There is a compelling need for additional research to ascertain
the conditions under which LLMs can be appropriately utilized in healthcare,
but also to establish conditions of gradual experimentation and learning
that align with principles of health ethics.

\section{Conclusion}

This review addresses ethical considerations of using LLMs in healthcare
at the current developmental stage. However, serval limitations are
import to acknowledge. Ethical examination of LLMs in healthcare is
still nascent and struggles to keep pace with rapid technical advancements.
Thus, the review offers a starting point for further discussions.
A significant portion of the source material originated from preprint
servers and did not undergo rigorous peer review, potentially introducing
limitations in their quality and generalizability. Additionally, the
findings\textquoteright{} generalizability may be limited due to variations
in researched settings, applications, and interpretations of LLMs.

\section*{Declarations}

\subsection*{Author Contributions}

JH and RR designed the protocol. JH conducted the search. Screening
data extraction and analysis was jointly conducted by JH and RR as
described in the methods section. JH wrote the first draft of the
manuscript which was, then, revised by RR. RR is the PI of the project
from which this article derives. All authors read and approved the
final version.

\subsection*{Declaration of Funding}

This study was funded by the VolkswagenStiftung as part of the Digital
Medical Ethics Network (grant number 9B233). The funder played no
role in study design, data collection, analysis and interpretation
of data, or the writing of this manuscript.

\subsection*{Conflict of Interest}

All authors declare no financial or non-financial competing interests.

\subsection*{Data Availability Statement}

The datasets used and/or analysed during the current study available
from the corresponding author on reasonable request.

\subsection*{Acknowledgement}

Fig.1 and Fig. 2 created with BioRender.com

\printbibliography

\newpage{}

\cleardoublepage{}

\appendices{}

\onecolumn

\section{Key concepts for the search}
\begin{enumerate}
\item LLM are machines which use computational methods, partially under
human supervision, to extract statistical relationships from large
corpora of data such as text documents, databases and websites. This
allows them to predict a number of tokens or words given a certain
input and thereby to generate text that follows the learned statistical
conventions to an extent that most humans find to mimick human language
use.
\item Healthcare is the practice of restoring, maintaining or improving
a physical or mental well-being, including prevention, diagnosis,
treatment or cure of disease or injury. Healthcare is often performed
in structured settings involving the help of trained health professionals
such as medical doctors, nurses etc.
\item Ethical~issues can be defined as a state of affairs in which moral
implications of a given situation cannot be determined without much
reservation, disagreement with regard to the right course of action
exists, or conflicting moral obligations presenting themselves. This
may include a variety of different situations, such as occurrence
of unclear or undetermined benefits, chances, risks or harms, unclear,
undetermined or conflicting views about addressees of moral complaints
as well as situations in which applicable moral principles obviously
should have been considered but weren\textquoteright t. We determined
that that an ethical issue exists whenever it is understood as such.
\end{enumerate}

\section{Overview on data extraction fields}
\begin{enumerate}
\item Bibliographic data {[}author; title; journal/server; funding; coi{]}
\item Article type {[}orientation;type{]}
\item Aims of publication
\item Field of application {[}name of device; device description; anticipated
development; technical intentionality\slash capabilities; technical
limitations; medical domain; applications and outcomes; benchmarks
for comparison; 
\item Value dimension {[}value; normative background; ethical methods{]}
\item Findings {[}user-interactions; concerns; perception of hype/fears;
systemic perspectives{]}
\item Recommendations
\item Implications for practice
\item Future research directions
\end{enumerate}

\end{document}